\documentstyle[preprint,epsf,aps,subeqnarray]{revtex}
\begin{document}
\draft

\def\strutdepth{\dp\strutbox}
\def\nw#1{\strut\vadjust{\kern-\strutdepth\vtop to0pt{\vss\hbox to\hsize
{\hskip\hsize\hskip5pt$\leftarrow$\hss\strut}}}{\em #1}}

\def\rmn{r_{m}}

\title{
 Coalescence of Liquid Drops
    }
\author{Jens Eggers$^*$, John R. Lister$^\dagger$, Howard A.
Stone$^\ddagger$}
\address{
$^*$Universit\"at Gesamthochschule Essen, Fachbereich Physik,
45117 Essen, Germany \\
$^\dagger$Department of Applied Mathematics and Theoretical Physics,
University of Cambridge, Silver St, Cambridge CB3 9EW, UK \\
$^\ddagger$Division of Engineering and Applied Sciences, Harvard
University, Cambridge, MA 02138, USA
    }

\maketitle
\centerline{9 March 1999}\vskip-24pt
\begin{abstract}
When two drops of radius $R$ touch, surface tension drives an initially
singular motion which joins them into a bigger drop with smaller surface
area.  This motion is always viscously dominated at early times.  We
focus on the early-time behavior of the radius $\rmn$ of the small bridge
between the two drops.  The flow is driven by a highly
curved meniscus of length $2\pi
\rmn$ and width $\Delta\ll\rmn$ around the bridge, from which we
conclude that the leading-order problem is asymptotically equivalent to
its two-dimensional counterpart. An exact two-dimensional solution  for
the case of inviscid surroundings  [Hopper, J. Fluid Mech. 
${\bf 213}$, 349 (1990)] shows that
$\Delta \propto \rmn^3$ and $\rmn \sim (t\gamma/\pi\eta)\ln
[t\gamma/(\eta R)]$; and thus the same is true in three
dimensions.  The case of coalescence with an external viscous fluid is
also studied in detail both analytically and numerically. A
significantly different structure is found in which the outer fluid forms
a toroidal bubble of radius
$\Delta \propto \rmn^{3/2}$ at the meniscus and $\rmn \sim
(t\gamma/4\pi\eta) \ln [t\gamma/(\eta R)]$. This basic difference
is due to the presence of the outer fluid viscosity, however 
small. With lengths scaled by
$R$ a full description of the asymptotic flow for $\rmn(t)\ll1$
involves matching of lengthscales of order
$\rmn^2$,$\rmn^{3/2}$, $\rmn$, $1$ and probably $\rmn^{7/4}$.

\end{abstract}

\section{Introduction}

Considerable interest has been devoted recently to the breakup of
free-surface flows into drops under the action of surface tension
(Rallison 1984; Stone 1994; Eggers 1997). Here we investigate the
complementary problem of surface-tension-driven coalescence of two
drops, which is of fundamental importance in understanding the
possible topological transitions in three-dimensional free-surface
flows.  For example, numerical implementations of merging (LaFaurie,
Nardone, Scardovelli, Zaleski \& Zanetti 1994) are based on
phenomenological prescriptions for joining the two surfaces, without a
fundamental understanding or description of the dynamics. 

Traditional applications of coalescence ideas include the
description of two-phase dispersions.  As an important
example, we mention phase separation in two-phase flows (Nikolayev,
Beysens \& Guenoun 1996; Bonnecaze, Martula \& Lloyd 1998), 
where the velocity field induced by
the merging of two drops entrains other drops, thus enhancing the rate
of coalescence. Another classical problem connected with drop
coalescence is sintering, i.e. the merging of a powder into a
homogeneous material by heating. In many cases, in particular that of
ceramics or glasses, bulk fluid motion is the dominant mechanism for
coalescence and the dynamical process is known as viscous
sintering.  In a classical paper, Frenkel (1945) posed the problem
of the merging of two spheres by slow fluid motion as the first step
towards understanding the properties of the material that results from
sintering.  For a different case in which surface diffusion is the
dominant mechanism of mass transport, the asymptotics of coalescence
has recently been worked out (Eggers 1998), but such surface-dominated
transport is very different from the bulk fluid motion of interest
here.

Much of the experimental and numerical work on coalescence in viscous
systems is motivated by the viscous sintering problem. An exception is
an experimental paper (Bradley \& Stow 1978) on the coalescence of
water drops, but the low viscosity of water makes the motion very
rapid and difficult to observe. On the other hand, by using a very
high viscosity fluid, the motion can be slowed down as much as desired
(Brinker \& Scherer 1990) and the experimental results agree very well
with numerical simulations of the Stokes equations (Martinez-Herrera
\& Derby 1995).  The only theoretical analysis of three-dimensional
coalescence is the qualitative work by Frenkel (1945).
Analytical solutions for two-dimensional coalescence (i.e. of parallel
cylinders) have been obtained using complex variable
techniques for the special case where the outer
fluid is perfectly inviscid or absent  (Hopper 1990, 1992, 1993a,b;
Richardson 1992).   We show below that the three-dimensional
problem has the same asymptotic behaviour as this two-dimensional
solution at early times. Our main aim, however, is to address the
more general case of coalescence with a viscous outer
fluid, for which we find that the structure of the solution near
coalescence is quite different from the case of an inviscid exterior,
though there is again a parallel between the two-dimensional and
three-dimensional problems. 

Part of the challenge in treating three-dimensional coalescence
arises from the fact that it starts from a singular initial
condition, shown in figure \ref{fig1}. We assume that the
drops are initially spherical, which is based on an underlying
assumption of negligible velocity of approach and hence negligible
hydrodynamic deformation before contact. We imagine that two such
drops have just been joined along their symmetry axis by some
microscopic mechanism to form a tiny bridge of radius
$\rmn$. Evidently, the ``meniscus'' around the bridge will be a region
of very high curvature, which drives the increase of $\rmn$
with time. Our main concerns will be the time dependence of
$\rmn$ for very early times, and the shape of the interface and
the flow field near the meniscus.  We note that the ratio of the
coefficient of surface tension
$\gamma$ and the viscosity
$\eta$ gives a fixed velocity scale $\gamma/\eta$, and thus the
expected Reynolds number
$Re = \rho\gamma\rmn/\eta^2$ will be be arbitrarily small as $\rmn\to0$,
and the flow will initially be described by the Stokes equations
regardless of the material parameters. 

In the next section, we set up an integral representation of the
Stokes flow  and split it into two parts: an outer region far
from the meniscus, in which the shape is still close to the
initial spherical condition, and an inner
region near the meniscus in which the shape evolves rapidly. The dominant
contribution to the velocity field comes from the high
curvature of the meniscus, and its amplitude is determined by the
lengthscale
$\Delta$ of this curvature.  The main task is thus to
find the structure and scale of this inner solution. Since the inner
solution is determined by the local curvature and not the global shape,
the results obtained here are not restricted to the simple
spherical geometry shown in figure
\ref{fig1}.  In section III we show that only a negligible fraction of the
fluid caught in the narrow gap between the two spheres is able to escape.
The rest accumulates in a toroidal pocket, or bubble, of radius
$r_b \propto \rmn^{3/2}$ that forms at the meniscus. This bubble is
connected to a thin neck of width $r_n \propto \rmn^2$. We have performed
extensive simulations for the simplest case of equal
viscosity fluids  which
confirm these scaling laws.

In section IV we examine the inner ``bubble'' solution
in greater detail. The bubble is joined to the neck by a short region of
very large curvature on a lengthscale that appears numerically to be
proportional to
$(r_br_n)^{1/2}$ and thus tends to a corner as $r_n/r_b\to 0$ (i.e.
$\rmn\to0$). Though the curvature of this corner is much greater than
that of the bubble, both contribute at the same order to the leading-order
motion of the meniscus, which can be thought of as simply due to a ring
force of strength
$2\gamma$ smeared over a lengthscale
$r_b$. In the final section we discuss the case of arbitrary
viscosity ratios and mention  related problems, namely,
the effect of arbitrary initial shapes and the scaling at zero
outer viscosity.

\section{From three to two dimensions}

For simplicity we consider two initially spherical drops of equal radii
$R$, as shown in figure \ref{fig1}. Simple
extensions to the  cases of unequal radii and non-spherical shapes will
be described in Section V. We denote the viscosity of the drops by
$\eta$ and that of the outer fluid by
$\lambda^{-1}\eta$.  As we have noted, the dynamics immediately after
coalescence is described by the Stokes equations. Since these equations
are linear, the velocity field can be expressed as an integral of the
driving surface forces
$\gamma \kappa {\bf n}$, where ${\bf n}$ is the normal directed into the
outer fluid and $\kappa=\nabla.{\bf n}$ is the curvature of the interface.
We make the velocity dimensionless with respect to  $\gamma/\eta$, all
lengths with respect to $R$, and times
with respect to the corresponding timescale $\tau = R\eta/\gamma$.

Calculation of the evolution of the interface $S(t)$ requires only the
interfacial velocity, which is given by the integral
equation (Rallison \& Acrivos 1978)
\begin{equation}
\label{int}
\frac{(1+\lambda^{-1})}{2}{\bf u}({\bf x}_1) = -\int_{S(t)} \kappa {\bf 
J.n}\,{\rm d}\sigma_2 + (1-\lambda^{-1})
\int_{S(t)} {\bf u.K.n} \,{\rm d}\sigma_2,
\end{equation}
where
\begin{equation}
\label{3d}
{\bf J(r)} = \frac{1}{8\pi}\left[\frac{{\bf I}}{r}
+ \frac{{\bf rr}}{r^3}\right], \qquad
{\bf K(r)} = -\frac{3}{4\pi} \frac{{\bf rrr}}{r^5}, \qquad
{\bf r} = {\bf x}_1 - {\bf x}_2, 
\end{equation}
$d\sigma_2$ denotes a surface area element at position ${\bf x}_2$, and
${\bf x}_1, {\bf x}_2$ both lie on
$S(t)$. The first term on the right-hand side of (\ref{int}) represents
the driving by the surface forces, while the second accounts for the
difference in viscosity between the fluids.  The
problem is closed by requiring that any material marker
$\xi$ with position ${\bf x}_1$ on the surface moves according to
\begin{equation}
\label{marker}
\partial_t {\bf x}_1(\xi) = {\bf u}({\bf x}_1).
\end{equation}

Equation (\ref{int}) and the identity 
\begin{equation}
\label{intJ}
\int_S{\bf J}.{\bf n}\, {\rm d}\sigma = {\bf 0},
\end{equation}
(which is a consequence of the incompressibility condition $\nabla.{\bf
J}={\bf 0}$) show that there would be no flow if $\kappa$ were constant over
$S$. It follows that, in the early stages of coalescence when $\rmn \ll
1$, the flow is driven by the small region around
the meniscus where $\kappa$ is not close to its initial constant
value of 2. This key observation motivates an analysis based on splitting
the interface into two regions.

We use cylindrical polar coordinates
$(r,z)$ with origin at the junction between the drops. 
Away from the
region of coalescence, the surface is essentially undisturbed and thus
has the form $h(z) = (2z)^{1/2}$ and $h(z) = (-2z)^{1/2}$ for $h\ll1$ in
$z>0$ and $z<0$, respectively. The width of the gap between the spheres
is given by
\begin{equation}
\label{w}
\phantom{\qquad\qquad(\rmn\ll r\ll1)}
w = r^2 \qquad\qquad(\rmn\ll r\ll1).
\end{equation}
and, since $\partial w/\partial r\ll1$, the interfaces on either side of
the gap are nearly parallel.

The solution for this outer region has to be matched with an
inner solution on the scale $r=\rmn$ of the bridge or meniscus where
the two drops are joined. The inner solution has a region of very
high curvature, which provides the dominant contribution to the velocity.
To a first approximation, this region can be represented as  a ring of
radius $\rmn$ and small width $\Delta$ connected to two asymptotically
straight interfaces each pulling outward with unit tension. The resultant
effect is that of a radially directed ring force with strength 2 per unit
length of the ring applied over a width $\Delta$.

To find the velocity field generated by
this ring, we try integrating over a circular line ${\cal L}$
of forces ${\bf f(r)} = 2{\bf e}_r$. Considering, for the moment, only 
the simple case $\lambda = 1$, for which
${\bf u}$ can be computed directly from (\ref{int}), we have
\begin{equation}
\label{line}
{\bf u}({\bf x}_1) = \frac{1}{8\pi} \int_{\cal L}\left[
\frac{{\bf f}({\bf x}_2)}{|{\bf x}_1-{\bf x}_2|} +
\frac{({\bf x}_1-{\bf x}_2)({\bf f}({\bf x}_2) . 
({\bf x}_1-{\bf x}_2))}{|{\bf x}_1-{\bf x}_2|^3}\right]
{\rm d}\sigma_2.
\end{equation}
From this representation it is evident that the force
distribution cannot be represented by a line everywhere as the
first term in the integral (\ref{line})
would lead to a logarithmically infinite value of the velocity.  
In the neighborhood of $r_m$, when $|{\bf x}_2-{\bf x}_1|=O(\Delta)$, one
must account for the fact that the force is distributed over a length
scale of size $\Delta$. The logarithmically dominant part of the
integral, which comes from $\Delta\ll |{\bf x}_2-{\bf x}_1|\ll\rmn$,
gives a radially directed flow
\begin{equation}
\label{log}
{\bf u}(\rmn) = -\frac{1}{2\pi} \ln\left(\frac{\Delta}{\rmn}\right)
{\bf e}_r .
\end{equation}

Since the curvature of the
ring is not apparent at leading order in the region $\Delta\ll |{\bf
x}_2-{\bf x}_1|\ll\rmn$ that dominates (\ref{log}), we may equivalently
consider the corresponding two-dimensional problem, in which coordinates
$(x,y)$ take the place of $(z,r)$. In that case (two parallel cylindrical
drops connected along a narrow band of width
$2\rmn$) the high-curvature meniscus is represented by two straight
lines a distance $2\rmn$ apart. Since the forces $2\gamma$ on the
lines pull in opposite directions, they cancel on scales much greater
than their distance apart, the integral (\ref{line}) is cut off on the
scale $\rmn$, and (\ref{log}) again results.
Because of this asymptotic equivalence of the two-dimensional and
axisymmetric problems, we will mostly
consider the two-dimensional problem from now on, which is simpler
numerically. The two-dimensional forms of the kernels ${\bf J}$ and
${\bf K}$ can be derived by integrating
(\ref{3d}) along the third dimension to obtain
\begin{equation}
\label{2d}
{\bf J(r)} = \frac{1}{4\pi}\left[-{\bf I}\ln r
+ \frac{{\bf rr}}{r^2}\right] ,\qquad{\bf K(r)} = -\frac{1}{\pi}
\frac{{\bf rrr}}{r^4},
\qquad {\bf r} = {\bf x}_1 - {\bf x}_2,
\end{equation}
and the surface integral (\ref{int}) is now along the 
perimeter of the two-dimensional drops.

Not only does (\ref{log}) give the leading-order velocity in both two and
three dimensions, but it also holds for all viscosity ratios $\lambda$.
A summary of the argument is as follows. It is clear that the early flow can
always be thought of as driven by ring or line forces of strength $2{\bf
e}_r$ or
$2{\bf e}_y$. Since Stokes flow has no inertia, this force is transmitted
unaltered across any surface enclosing the bubble and, since the width of
the gap between the spheres is asymptotically negligible, the force must be
supported by the internal fluid, and the external fluid in the gap makes
little difference. Now the logarithmically large velocity of a slender body
moving under a given force does not depend on the viscosity of the body, as
can be seen explicitly in the solution for the motion of a cylinder of one
fluid through another fluid (Lister \& Kerr 1989). Applying
this result to the bubble at the meniscus,
the finite viscosity of the external fluid in the bubble also makes little
difference and  (\ref{log}) is correct at leading order for
$\lambda\ne1$ (though the higher-order corrections do depend on $\lambda$). 

To evaluate the velocity from (\ref{log}), however, it is necessary to
determine the scale
$\Delta$ over which the force is distributed at the meniscus and here the
viscosity ratio does play a role. In the following section we will show
that
$\Delta\propto r_m^{\alpha}$, where 
$\alpha=3/2$ for finite $\lambda$ and $\alpha=3$ for the
special case $\lambda=\infty$ (no outer fluid). By integrating
(\ref{log}), we find that
\begin{equation}
\label{tlog}
r_m(t) \sim -\frac{(\alpha-1)}{2\pi}t \ln t.
\end{equation}
Recalling that time is measured in units of $\eta R/\gamma$, we see
that the estimate based on dimensional analysis alone, $r_m \propto t
\gamma/\eta$ is not quite correct, and in fact requires a logarithmic
correction.

\section{Asymptotic shape of the meniscus}

Using the equivalence of the two- and three-dimensional problems, we
now study the coalescence of two viscous circular cylinders in more
detail. Figure \ref{fig1a} compares Hopper's exact solution for
$\lambda =
\infty$ with a numerical simulation  for $\lambda = 1$ in which the
initial condition is that of a cusp with $r_m = 10^{-3}$, smoothed on the
scale of $r_m^2$. (The numerical method is described in more
detail below.)  The shape of the
meniscus for $\lambda=\infty$ is the tip of a near cusp (Fig.
\ref{fig1a}a), while the shape for $\lambda = 1$ is observed to be 
quite different for most of the evolution  (Fig.
\ref{fig1a}b): the external fluid is collected in a small bubble
at the meniscus, making the lengthscale $\Delta$ of the local
solution much larger than in the absence of an external fluid.  Only in the
last stages of merging does the fluid caught inside the bubble 
escape and the results look qualitatively more like Hopper's solution
(Fig. \ref{fig1a}c).

\subsection{Analysis for an inviscid exterior}

The existence
of an exact two-dimensional solution for the
special case $\lambda=\infty$ of an inviscid or absent external
fluid (Hopper 1993a,b; Richardson 1992) allows us to test the general
ideas of section II. Asymptotic expansion of
this solution near the meniscus shows that
\begin{equation}
h(x) \sim \left[\textstyle{1\over 2}\rmn^2 + \sqrt{ \left(\textstyle{1\over
2}\rmn^2\right)^2
 + (2x)^2}\right]^{1/2} \qquad\qquad(x\ll1),
\end{equation}
from which we deduce that the highly curved region is of size
$\Delta \sim \rmn^3$, and the curvature $\kappa$ scales like $\Delta^{-1}
\sim
\rmn^{-3}$. This result is somewhat surprising since $\Delta$ is much
smaller than the gap width $w \sim \rmn^2$ estimated from the spherical
shape of the outer solution and, at present, we do not have an asymptotic
argument for the appearance of the small scale $\rmn^3$. (By contrast,
with a  viscous outer fluid 
$\Delta\sim \rmn^{3/2}$ is {\it larger} than the gap width $w$, and this
scale can be understood from mass conservation as discussed in 
section III B.)

Inserting $\Delta\sim \rmn^3$ into (\ref{log}),
we obtain
\begin{equation}
v \sim -\frac{1}{\pi} \ln \rmn 
\end{equation}
for the velocity ($v=\dot{r}_m$) at the meniscus, which agrees with
the asymptotic result given by Hopper (1993a); our earlier
asymptotic analysis  now allows the extension of this result to three
dimensions. 

\subsection{Analysis for a viscous exterior}

For the case $\lambda =1$ we begin by discussing 
the structure of the local solution
close to the meniscus, which is shown in figure \ref{fig2}.
It consists of a ``bubble'' of outer fluid of radius $\Delta\equiv r_b$,
which is connected to a thin neck of width $r_n$. The neck
matches onto the static outer solution, so $r_n$ must
scale like the gap width $w \sim \rmn^2$. The area of 
the original gap up to $r_m$ is $O(r_m^3)$
so that, if the meniscus advances faster than fluid 
can escape from the gap, the bubble should contain
a finite fraction of the gap fluid and hence $r_b \propto r_m^{3/2}$.

To examine this argument in more detail, we consider the velocity field
generated by the large curvature of the meniscus. As we have already
noted, the flow is driven by that part of the interface  where the
interfacial curvature is significantly different from 1 (or 2 in the
spherical case), namely 
$|r-\rmn|=O(r_b)$. Using a multipole expansion of this forcing,
we find the velocity field ${\bf u}({\bf x}_1)$ at a distance 
$|{\bf x}_1 - {\bf x}_m| \gg r_b$ from the center ${\bf x}_m$ 
of the bubble to be 
\begin{equation}
\label{lead}
{\bf u}({\bf x}_1) = {\bf f}.{\bf J}({\bf x}_1-{\bf x}_m) ,
\end{equation}
at leading order, where ${\bf J(r)}$ is given by (\ref{2d})
and ${\bf f}$ is the total force exerted by the bubble.
This force is the integral of $-\kappa {\bf n} = \partial_s {\bf t}$ 
over the bubbles' surface (${\cal L}_b$), where ${\bf t}$ is
the tangent vector pointing in the direction of increasing 
arclength $s$: 
\begin{equation}
\label{force}
{\bf f} = -\int_{{\cal L}_b} \kappa {\bf n} {\rm d}\sigma = 
\int_{{\cal L}_b} \partial_s {\bf t} {\rm d}\sigma = 
{\bf t}_2 - {\bf t}_1.
\end{equation}
In the present case $-{\bf t}_1 = {\bf t}_2 = {\bf e}_y$, so that 
the far-field velocity resulting from the forcing of the bubble and 
its image is
\begin{equation}
\label{multi}
{\bf u}({\bf x}_1)=2[{\bf J}({\bf x}_1-{\bf x}_m)-
{\bf J}({\bf x}_1+{\bf x}_m)].{\bf e}_y
\end{equation}

We are interested in the flow in the neck, so we choose ${\bf
x}_1=(0,y_1)$  with
$y_1-\rmn\gg r_b$. From (\ref{2d}) and (\ref{multi}) we find that 
the $y$-component of velocity in the neck is given by
\begin{equation}
\label{un}
u_n(y_1) = -\frac{1}{2\pi}\ln\left(\frac{y_1-\rmn}{y_1+\rmn}\right)
 .
\end{equation}
The representation (\ref{un}) breaks down when $y_1-\rmn=O(r_b)$, since
the higher-order terms in the multipole expansion become of
comparable magnitude and $u_n(y_1)$ crosses over to some function
that depends on the detailed structure of the bubble. Since (\ref{un})
must match onto the velocity field in the bubble, we can write the
velocity of the meniscus as
\begin{equation}
\label{v}
v = v_0 + \frac{1}{2\pi}\ln\left(\frac{2\rmn}{r_b}\right),
\end{equation}
where the constant
$v_0$ comes from the detailed shape of the bubble, and for 
$\lambda = 1$ is found numerically to be
$v_0 = -0.077$. (This equation is consistent with (\ref{log}), but
also includes a representation of the $O(1)$ contribution.) 

If $r_b\ll y_1 - \rmn \ll \rmn$ then from 
(\ref{un}) and (\ref{v})
\begin{equation}
\label{una}
u_n(y_1)\approx
\frac{1}{2\pi}\ln\left(\frac{2\rmn}{y_1-\rmn}\right)\ll
\frac{1}{2\pi}\ln\left(\frac{2\rmn}{r_b}\right)  \approx v.
\end{equation}
Thus the fluid 
in the neck ahead of the bubble only starts to move at a speed
comparable to that of the bubble when $y_1 - \rmn=O(r_b)$ i.e. when the
bubble has caught up with it. Thus all the fluid in the neck is collected
into the advancing and growing bubble.
Since the neck width $r_n$ scales like $\rmn^2$, the total area of neck
collected scales like
$\sim r_m^3$, and thus
\begin{equation}
\label{vol}
\quad r_b \sim \rmn^{3/2} .
\end{equation}

Finally, combining (\ref{vol}) with (\ref{v}) or  (\ref{tlog}), we find
that
\begin{equation}
\label{rd}
v \sim -\frac{1}{4\pi} \ln \rmn ,
\end{equation}
which can be integrated to give
\begin{equation}
\label{r}
\rmn \sim -\frac{t}{4\pi} \ln t .
\end{equation}
This result differs by a factor of $4$ from the $\lambda = \infty$
limit.

\subsection{Numerical tests of scaling}

To test the predictions of these scaling ideas, we have performed
extensive two-dimensional simulations of drop coalescence using
the boundary-integral method. The initial bridge has
$r_m(0) = 10^{-6}$, which means
the gap width is $w = 10^{-12}$ initially. For simplicity,
we only considered the viscosity matched case $\lambda =1$ so that the
second term drops out from (\ref{int}) and
${\bf u}$ can be computed directly from the surface forces.
The interface was parameterized by arclength, and derivatives
were evaluated using centered differences. The interface was
advanced according to (\ref{marker}), using an explicit second-order 
Runge-Kutta step. The difference between the result
of time step $\Delta t$ and two half-steps $\Delta t/2$
was used to control the time step. 

Improvement of the stability
of our numerical method by making it 
implicit would be
computationally very demanding for an integral operator. Instead, we use a
scheme first proposed by Douglas \& Dupont (1971). The equation of
motion (\ref{marker}) for the position of the interface
can be written at any given time as a linearization around 
the current interfacial position ${\bf x}_0$:
\begin{equation}
\partial_t {\bf x} = {\bf A}({\bf x})\cdot({\bf x}-{\bf x}_0) + constant
\end{equation}
By writing this as 
\begin{equation}
\partial_t \delta{\bf x} = ({\bf A-B})\delta{\bf x} + 
{\bf B}\delta{\bf x} + constant
\end{equation}
and treating the first part explicitly, but the second part
implicitly, the scheme becomes unconditionally stable as long 
as $|{\bf B}| > |{\bf A}|/2$, where the matrix norm is defined to be the
modulus of the largest eigenvalue (Douglas
\& Dupont 1971).
In the present case,  
$|{\bf A}|$ scales like
$(\Delta x)^{-1}$ up to logarithmic corrections, where $\Delta x$ is the
minimum grid spacing.  By choosing ${\bf B}$ to be a diffusion operator
multiplied by  the local grid spacing, one can make sure that the
numerical  method, although treating the integral operator explicitly, 
becomes unconditionally stable. Without the help of this
trick the time steps required 
to integrate over the first decade and a half in $r_m$ would
have been prohibitively small.

To achieve the necessary spatial resolution, local refinement of the
mesh is crucial. The resolution near the meniscus was set by
the inverse of the local curvature. Away from the
inner solution, the grid spacing was allowed to taper off
geometrically, with the spacing constrained to change by no
more than 10\% from one grid point to the next. As explained
in more detail below, additional resolution was used in the
transition region where the bubble merges into the neck. The
maximum number of points used to represent one quadrant of
the shape was about 900. Every few time steps a new grid
was constructed using the current interface, and the interface
was interpolated to the new grid. Thus there was no need to
rearrange the grid points along the interface.

In figure \ref{fig3} we show the scaling of the bubble radius $r_b$
and the minimum neck radius $r_n$ and observe that both follow the
predicted power laws (\ref{vol}). A closer inspection shows that the
slope of $\log r_b$ is slightly smaller than expected, which is because
the scaling of the area of the bubble is almost the same as that of the
neck. (The portion of the neck up to the
point $y = \rmn + a r_b$ contributes an area
$A_n \approx a r_b \rmn^2 \sim \rmn^{7/2}$, which is only slightly
smaller than $A_b\approx \rmn^3$ since the two exponents are close.)
We confirmed numerically that the $r_m$-dependence of the total
area $A = A_b+A_n$ has no significant deviation from $r_m^3$, as
expected.  

The scaling of the velocity at the meniscus ($v$) and in the
neck ($u_n$) at position $y_1 = \rmn + a r_b$ is shown in figure
\ref{fig4} (with
$a=20$) and compared with $(1/2\pi)\ln(2\rmn/r_b)$, which is predicted by
equations (\ref{v}) and (\ref{una}) to have the same slope.  The
theory gives $v-u_n$= constant = $\ln a/(2\pi) + v_0$ which is
found numerically to be very close to $0.4$ and hence
$v_0 = -0.077$.  All three curves in figure
\ref{fig4} should have a slope $\ln 10/4\pi$ when plotted against
$\log_{10}(\rmn)$.  The noticeable deviation comes from the fact that one
is effectively taking the {\it difference} between $\log_{10} \rmn^{3/2}$
and
$\log_{10} \rmn$, so non-asymptotic effects in $r_b$, still present on
these small scales, become more pronounced.

It has been implicit in the previous arguments that the inner
solution, consisting of a bubble connected to a thin neck,
has reached its asymptotic form: in the frame of reference
of the advancing bubble tip, and rescaled by the bubble radius,
the shape should be stationary. Figure \ref{fig5}, showing both the local
interface profile and the curvature at two values of $r_m$ one decade apart,
reveals that there is, in fact, a slow variation in 
part of the local profile. This slow variation is seen as a 
positive second peak
in the curvature at the point where the bubble meets the neck.
To be able to resolve scales down to $\rmn=10^{-6}$, additional grid
points were inserted at the position of the second peak, where the
grid spacing was based on the width of the peak.
As $r_m\to 0$, which corresponds to going back in time, the
second curvature peak increases and also gets narrower, as its integral must
be finite to yield a finite change of slope. 

To obtain more information about the asymptotic shape of the
inner solution and the growth of the secondary peak in curvature,
it is useful to consider the inner solution as a separate
problem.
This analysis is given in the next section.

\section{Bubble on a neck}

Here we study the local solution close to the meniscus, which consists
of a bubble of radius $r_b$ connected to a thin neck of width $r_n$
(see inset to figure \ref{fig7}).  Asymptotically, the curvature of
the neck is very small compared with $r_b^{-1}$, so the
neck can effectively be considered as an infinitely long channel of
uniform width. From now on, all lengths will be measured in units of
$r_b$, so the radius of the bubble is normalized to unity, and the radius
of the neck asymptotes to some small number $\epsilon \approx r_n/r_b$. 
The solution we are interested in, which corresponds to the
asymptotic structure of the main solution described in Section III, is
such that the interfacial shape $g(y,t)$ is advected at a constant speed
$v_c$ without changing its shape
\begin{equation}
\label{conv}
g(y,t) = G(y - v_c t) .
\end{equation}
The physical meaning of this statement is that in the original
problem the local solution relaxes to a quasi-steady shape on a much
shorter timescale than the position $\rmn$ of the meniscus is changing.
The velocity field of the local  steadily translating shape  (\ref{conv})
must satisfy
\begin{equation}
\label{stat}
(u_y - v_c) G' = u_x .
\end{equation}
The boundary condition is that $g(y)$ approaches $\epsilon$ as $y
\rightarrow \infty$. The components of the velocity field $u_x$ and
$u_y$ follow from the integral equation (\ref{int}) as usual.  Instead
of solving the system (\ref{int}) and (\ref{stat}) directly, we found
it most convenient to evolve two bubbles attached to a very long,
straight neck of radius $\epsilon_{init}$ until a stationary shape is
established, as shown in the inset to figure \ref{fig7}.  The tension
in the neck is responsible for pulling the bubble along. The neck
shortens during the relaxation and the radius increases to a value
$\epsilon$, which then only changes very slowly by the time a
stationary shape is reached. 

Since the radius of the bubble $r_b$ and the radius of the neck $r_n$
are very different, one might think that the curvature distribution
within the bubble is independent of the neck radius
$\epsilon$. However, the limit $\epsilon \rightarrow 0$ turns out to
be singular as an increasingly pronounced peak of positive curvature
appears at the junction between bubble and neck, as demonstrated in
figure \ref{fig5}.

First, in figure \ref{fig7} we compare the curvature distribution as
given in figure \ref{fig5} for $r_m=10^{-3.5}$ with that of the
stationary problem, equations (\ref{int}) and (\ref{stat}), with
$\epsilon = r_n/r_b$. We choose $\epsilon_{init}$ such that the neck
has the appropriate width $\epsilon$ by the time a stationary shape is
reached. The excellent agreement shows that the flow close to the
meniscus is completely equivalent to the translating bubble, which is
of course a much simpler problem. Hence the inner solution of the
coalescence problem can be understood completely in terms of the
translating bubble.

Figure \ref{fig8}(a) shows a sequence of bubble shapes for
increasingly small values of $\epsilon$. The overall shape of the
bubble does not depend very much on $\epsilon$, and it looks as if the
shapes are almost the same for the two smallest values. However there
is an increasingly sharp ``corner''  at the point where the
neck meets the bubble. This result is most evident from a plot of the
curvature for the same values of $\epsilon$; see figure \ref{fig8}(b).
While the curvature of the bubble is negative, the corner at the
junction between the bubble and the neck corresponds to a growing peak
of positive curvature. In figure \ref{fig8}(b) we also include a plot
of the maximum of this peak as a function of
$\epsilon$. 
The data is suggestive of
$\kappa^{-1}\sim\epsilon^{1/2}$ though it is hard to make a definite
statement. If this is so then the lengthscale of the corner is the
geometric mean of the scales of the bubble and the neck, which would be
$\rmn^{7/4}$ in the coalescence problem. From (\ref{force}) it 
is evident that the total force
exerted by the peak is equal to the change in slope. As can be seen 
from figure \ref{fig8} (a) the change is constant to a good approximation
so 
that, of the total force $2{\bf e}_y$ exerted by the bubble, roughly 15\%
is exerted by the corner and 85\% by the rest of the bubble.


\section{Discussion and Outlook}

As we have demonstrated, even the simplest case of viscosity-matched
fluids represents a problem of enormous complexity, in which there are
features on
 at least the lengthscales $\rmn^2$, $\rmn^{3/2}$, $\rmn$ and 1.
Therefore, we confined ourselves to computing the leading-order
asymptotics, which we note are only logarithmically dominant. In all
likelihood, quantities like the bubble radius
$r_b(t)$ contain additional logarithmic terms, whose calculation require
a better knowledge of the matched asymptotics of the problem. Further
complications arise because the inner solution is itself singular, with
a corner of lengthscale which we estimate to be $O(\rmn^{7/4})$. 
Important goals for future work would be confirmation of the inner
scalings, formal asymptotic matching of the different scales, and to go
beyond the leading-order problem.

Another important problem is the generalization of our calculations to
arbitrary viscosity ratios $\lambda$.  A major obstacle to developing a
quantitative theory for general $\lambda$ is that numerical simulations
are much more difficult. When $\lambda \ne 1$ (\ref{int})
is a second-kind integral equation for $\bf u$,
which requires an order $N^3$ effort to solve instead of $N^2$.
Moreover, for small $\rmn$ the matrix associated with the second
integral becomes singular since the local solution for
the velocity field is close to a uniform translation for
which the kernel has a zero eigenvalue. We have not yet found a way
to treat this singularity sufficiently well to go beyond
$\rmn$-values of $10^{-2}$ for $\lambda \ne 1$.

We have argued already  that we expect the leading-order behaviour of
$\dot\rmn$ to be given by (\ref{log}) for any $\lambda$ since the net
force from the meniscus is supported asymptotically by the internal fluid.
We also believe that the scaling $\Delta\sim\rmn^{3/2}$ will hold as
$\rmn\to0$ for any finite $\lambda$ since, even for large
$\lambda$ (small external viscosity), the pressure drop along a
narrow channel is large and so the outer fluid is not able to escape
from the gap and is caught in a bubble. Expressed differently, the positive
curvature at the bubble corner is unable to pull the walls of the channel
apart since this would require significant motion along the channel. It
is intriguing that Hopper's exact solution shows that the
situation is very different for
$\lambda = \infty$ when there is no external viscous resistance to overcome
and no bubble forms (figure 2a). Thus the limits $\lambda \rightarrow
\infty$ and $\rmn\to0$ do not commute, reflecting the fact that one must
be careful in assuming a zero-stress condition in situations involving
narrow cusps, as even a very small external viscosity can be significant.

We can provide a physical estimate for the scale below which a bubble 
forms for $\lambda
\gg 1$.  Fluid motion in the narrow gap may be treated with the lubrication
approximation, whence
$\eta u_n / (\lambda r_n^2) \approx \Delta p/r_m$ and $\Delta p\approx
\gamma/r_b$ gives the estimate $u_n
\approx (\lambda \gamma/\eta) (r_m/R)^{3/2}$. Therefore, a bubble can only
be expected to form when the meniscus motion $v \approx \gamma/\eta > u_n$,
which occurs on an approximate lengthscale $r_m /R < \lambda^{-2/3}$. 

We have been considering the simplest case of equal spheres brought
into contact. The case of unequal spheres of radii $R$ and $R/\delta$,
with
$\delta<1$ is a straightforward generalization in which the gap thickness
(\ref{w}) is simply replaced by
\begin{equation}
\label{w2}
\hskip2cm w = r^2(1 + \delta)/2 \qquad\qquad(\rmn\ll r\ll1).
\end{equation} 
Similar considerations show that the initial evolution of axisymmetric
drops brought into contact along their symmetry axes depends only on the
local curvature at the initial contact. An more interesting variation is
that of general initial shapes for which the locus of high
curvature (in three dimensions) near contact no longer forms a circle, but
is a more general closed curve. To leading order, this curve is convected
with a logarithmically large velocity field, pointing in the direction
normal to the curve.

Recent research (Nikolayev {\it et al.} 1996;
Bonnecaze {\it et al.} 1998) has suggested that the rate
of coalescence in emulsions can be greatly enhanced by the flow generated
by individual coalescence events. Our analysis  suggests that an
appropriate model for the far-field velocity of a single coalescence
event is the Stokes flow driven by an expanding ring force of radius
$r_m$ and strength $2\gamma$ per unit length. This dipolar flow, which may
be obtained by solving 
\begin{equation}
\label{ring}
\eta\nabla^2{\bf u} - \nabla p + 2\gamma
\delta(r-r_m)\delta(z){\bf e}_r = {\bf 0}
\end{equation}
using Hankel transforms \cite{SB}, is
\begin{subeqnarray}
u_r(r,z) &=& \frac{\gamma r_m}{2\eta}\int_0^{\infty}
J_1(kr)J_1(kr_m)(1-kz)e^{-kz} {\rm d}k \nonumber \\ 
u_z(r,z) &=& -\frac{\gamma r_m}{2\eta}\int_0^{\infty}
J_0(kr)J_1(kr_m)kze^{-kz} {\rm d}k . \\
p(r,z) &=& 2\eta u_z(r,z)/z \nonumber,
\end{subeqnarray}
where the $J_n$ are Bessel functions.
These equations can be approximated in the limit
$r_m \ll(r^2+z^2)^{1/2}$ to obtain the axisymmetric dipole
\begin{equation}
\label{rapprox}
{\bf u}(r,z) \approx \frac{\gamma r_m^2}{4\eta}
\frac{(r^2-2z^2)}{(r^2+z^2)^{5/2}}\left[r{\bf e}_r+z{\bf e}_z\right] ,
\end{equation}
which is shown in figure \ref{fig9} superimposed on the outlines
of two spherical bubbles. The coalescence-induced
radially directed force drives a flow towards the 
two drops  over an angle $\approx 127^\circ$. The flow (\ref{rapprox})
might be used for simplicity in models of multiple coalescence. 

All motions described so far begin with a local point contact and it
is worth considering how this contact might be achieved.  The near-contact
squeezing motion generated when two drops (or a drop and a plane) are in
relative motion can be analyzed using the lubrication approximation (e.g.
Jones \& Wilson 1978; Yiantsos \& Davis 1991).  Owing to the large pressures
accompanying flow along the narrow gap, the surface tends to deform in the
narrow gap.  In particular, when two equal size drops are squeezed together
with a force
$F$ on each drop, then a dimple tends to form when
$h_0(t)<R/\lambda^2$, where $h_0(t)$ is the
minimum gap spacing and $R$ the radius of curvature of the undeformed drop.
The magnitude of the deformation becomes the same order of magnitude as the
gap height and the dimple has a radial scale $\left( F R/\gamma
\right)^{1/2}$. Away from the gap the drop is nearly spherical so long as,
for $\lambda=O(1)$, the effective capillary number is small, $O( F
h_0^{1/2}/\gamma R^{3/2} ) \ll 1 $. An implication of these analyses
is that contact is very likely to occur along a rim, or at least at an
off-axis position, with an initial radius of the bridge of order $\left ( F
R/\gamma\right )^{1/2}$. The dynamics would then follow the results shown in
Section III.

From an experimental point of view, it is probably not relevant to
investigate smaller (dimensionless) $\rmn$ than $10^{-5}$ since the
gap width just ahead of the bubble is proportional to $\rmn^2$, which is
then of microscopic size for reasonable values of $R$. Very small
inhomogeneities in the fluid or van der Waals attractions will cause
the two interfaces to reconnect, and to create an instability that
breaks the azimuthal symmetry we have assumed in the
three-dimensional problem. Moreover, the bubble actually forms a
structure that resembles a long thin torus in three dimensions,
and is thus prone to a Rayleigh capillary instability, which grows on
a short timescale proportional to $r_b$ and is potentially dangerous.
On the other hand, there are stabilizing effects since the bubble is
also convected (Brenner, Shi \& Nagel 1994), and a careful nonlinear
stability analysis has to be done to determine the scale where the
stability is first expected to occur. This question is experimentally
relevant because it sets the size of small bubbles of the outer fluid
that may be observed after coalescence.

We have pointed out that coalescence is initially described by
the Stokes equations. If the viscosity of the fluid is small, this
is true only for the early stages of coalescence, until
the Reynolds number is of order one, which happens when
\begin{equation}
\label{ln}
\rmn \approx \ell,\quad\ell = \eta^2/(\rho\gamma) .
\end{equation}
For water, $\ell = 1.4 \times 10^{-6} \,\rm cm$ , so it is a very relevant
question to go beyond the Stokes approximation.  After passing the
transition region (\ref{ln}) we expect the dynamics to be described by
the Euler equations.  Assuming that the scale of the local solution at
the meniscus is set by the gap width alone,
the interfacial stress driving this motion is
approximately $\gamma/(r_m^2/R)$, which is to be the same magnitude as
$\rho v^2$. Hence, with $v = {\dot r}_m$, we find
\begin{equation}
\rmn \propto \left ( {\frac{\gamma R}{\rho }} \right )^{1/4} t^{1/2},
\end{equation}
which corresponds to $v \propto t^{-1/2}$. The
geometrical part of both problems is similar to the Stokes case, but the
coupling between pressure and velocity makes the relationship between the
surface shape and the velocity different, so an alternative treatment is
needed to predict the numerical coefficient in front of the power law for
$\rmn$. 

Clearly, the possibility of finite Reynolds number, arbitrary
surface shape, finite velocity of approach, and the inclusion of
another fluid outside the drops lends a tremendous richness to the
class of singularities studied here.

\acknowledgements
We are indebted to Stephane Zaleski for bringing the authors
together and for sharing his deep insight. J.E. and J.R.L.
were visitors of LMM at the University of Paris VI,
and H.A.S. of PCT at ESPCI. Denis Gueyffier helped us crucially
by producing the first simulation of coalescing drops. Todd Dupont
gave very helpful advice on the numerical simulations.

\begin{figure}
\caption{
The surface profile $h(z)$ produced by two coalescing drops of
radius $R$. The origin of the axis of symmetry
$z = 0$ lies at the initial point of contact. The bridge joining
the two spheres has radius $r_m$ and width $\Delta$.
}
\label{fig1}
\end{figure}

\begin{figure}
\caption{
A closeup of the point of contact during coalescence of two identical 
cylinders
for the two cases of no outer fluid ($\lambda = \infty$) and two fluids
of equal viscosity ($\lambda = 1$). 
(a) is Hopper's solution for $r_m = 10^{-3}, 10^{-2.5},
10^{-2}$, and $10^{-1.5}$. (b) a numerical simulation of the viscosity
matched case that shows fluid collecting in a bubble at the meniscus.
Note that the two axes are
scaled differently, so the bubble is almost circular.
For large values of $r_m$, as shown in (c), the fluid finally
escapes from the bubble, and the width of the meniscus is closer
to the value of the gap width $O(r_m^2)$.
}
\label{fig1a}
\end{figure}

\begin{figure}
\caption{
The structure of the local solution close to the meniscus.
It resembles a bubble connected to a thin neck. The radius
of the bubble is $r_b$ and the minimum radius of the neck is
$r_n$. The distance from the origin to the front of the bubble
is $r_m$, which is not drawn to scale here.
}
\label{fig2}
\end{figure}

\begin{figure}
\caption{
Scaling of the bubble radius $r_b$ and the neck radius $r_n$
as function of $r_m$.
}
\label{fig3}
\end{figure}

\begin{figure}
\caption{
Scaling of the velocity $v$ at the tip and the velocity $u_n$
at a position $y = r_m + 20r_b$ in the neck. There is
a constant difference of 0.4 between the two. Both agree
very well with the scaling of $(1/2\pi)\ln(2 r_m/r_b)$ as
predicted by theory. For comparison, we also give a
slope of $(\ln 10)/4\pi = 0.183$.
}
\label{fig4}
\end{figure}

\begin{figure}
\caption{
The local solution and its curvature distribution in
coordinates rescaled by $r_b$. As $r_m$ gets smaller, a sharp
peak develops at the junction between the bubble and
the neck.
}
\label{fig5}
\end{figure}

\begin{figure}
\caption{
A comparison of the curvature distribution of the local
solution for $r_m = 10^{-3.5}$ and the stationary ``bubble on
a neck''. The peak height of the positive curvature has
been used to match the two. In an inset, we show the initial
condition used to compute the stationary shape of
the translating bubble.
}
\label{fig7}
\end{figure}

\begin{figure}
\caption{
(a) The stationary state of a translating bubble, pulled by a
thin neck for different values of $-\log_{10}\epsilon = 1.8, 2.8,
3.7$, and $4.9$. (b) Blowup of the 
curvature distribution. Although the surface shape seems 
to have converged, the maximum negative curvature in the 
junction still increases. The insets show the scaling of
the maximum curvature $\kappa_{max}$ and that of $\kappa_{max}*w_{1/2}$,
where $w_{1/2}$ is the half-width measured in arclength.
}
\label{fig8}
\end{figure}

\begin{figure}
\caption{
The velocity field generated by an expanding ring of forces in the 
limit of the radius of the ring $r_m$ going to zero. Note 
that the velocity field is pointing outward in the direction
of the expanding ring, but is inwardly directed over
much of the flow domain.
}
\label{fig9}
\end{figure}


\newpage
\begin{figure}
  {\Huge Figure 1:}
  \begin{center}
    \leavevmode
    \epsfsize=0.9 \textwidth
    \epsffile{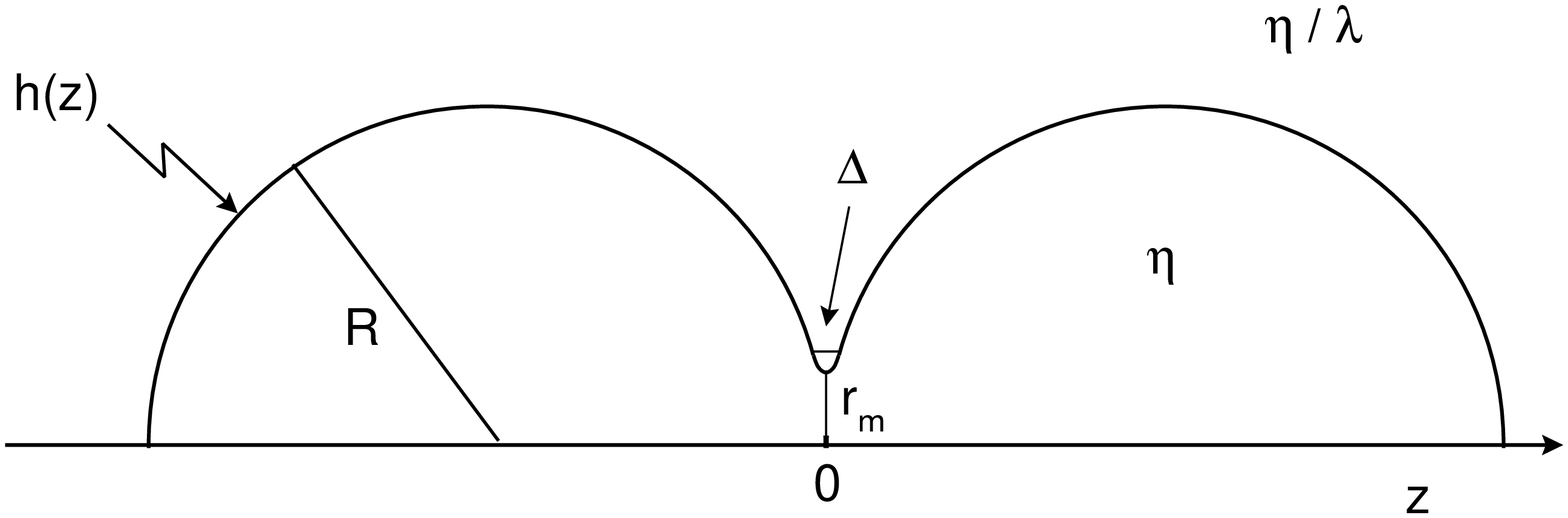}
  \end{center}
\end{figure}

\newpage
\begin{figure}
  {\Huge Figure 2a:}
  \begin{center}
    \leavevmode
    \epsfsize=0.9 \textwidth
    \epsffile{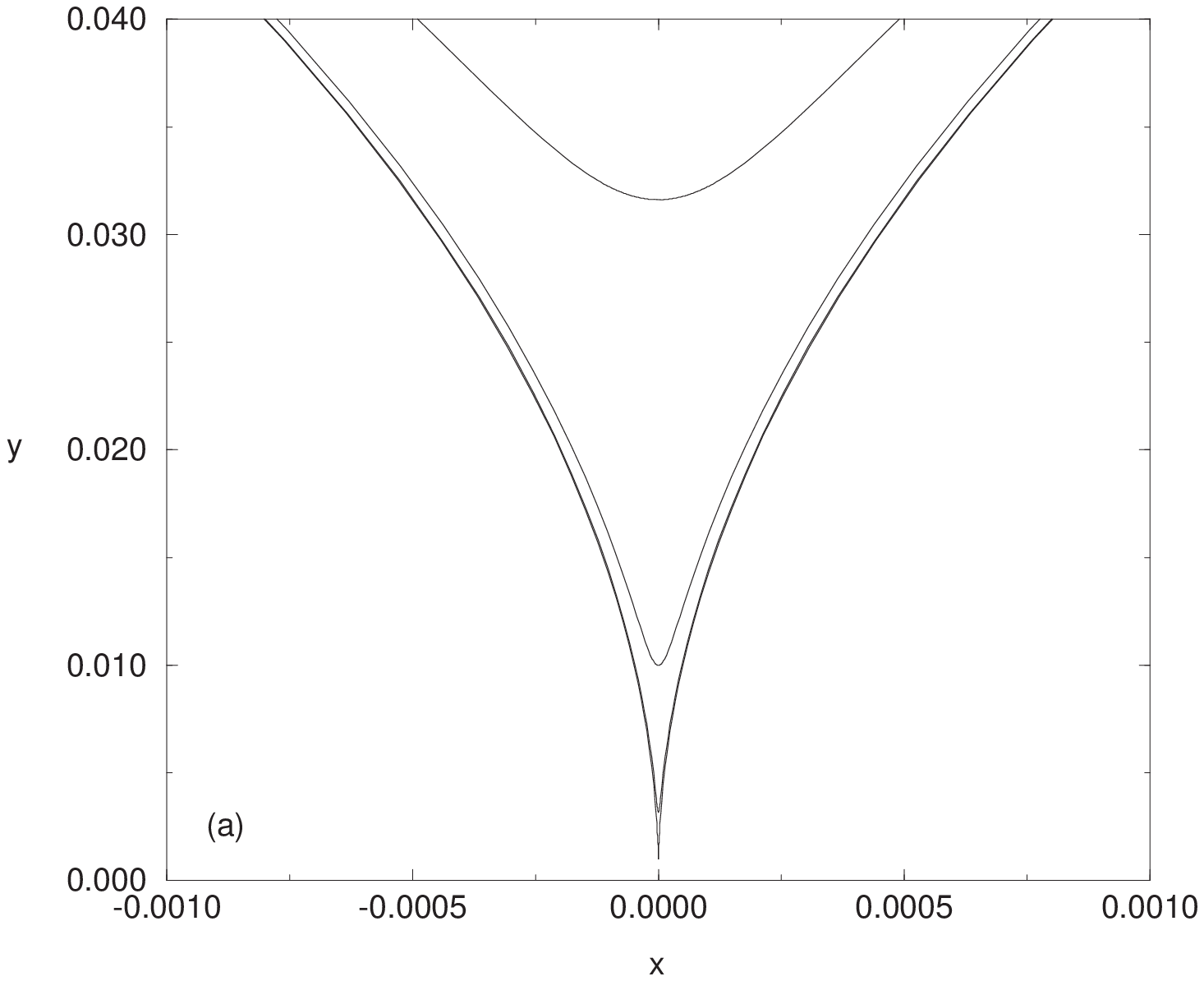}
  \end{center}
\end{figure}

\newpage
\begin{figure}
  {\Huge Figure 2b:}
  \begin{center}
    \leavevmode
    \epsfsize=0.9 \textwidth
    \epsffile{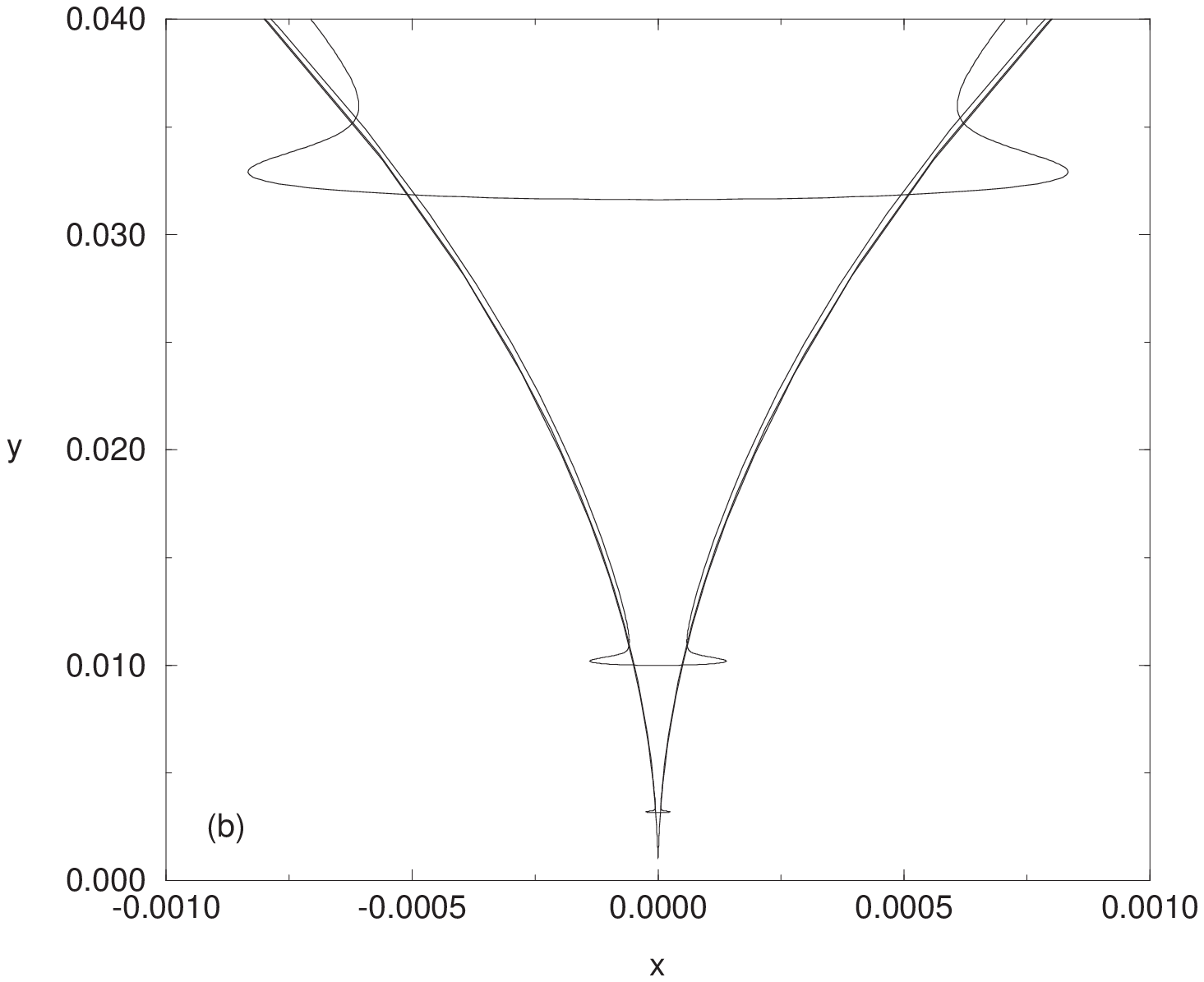}
  \end{center}
\end{figure}

\newpage
\begin{figure}
  {\Huge Figure 2c:}
  \begin{center}
    \leavevmode
    \epsfsize=0.9 \textwidth
    \epsffile{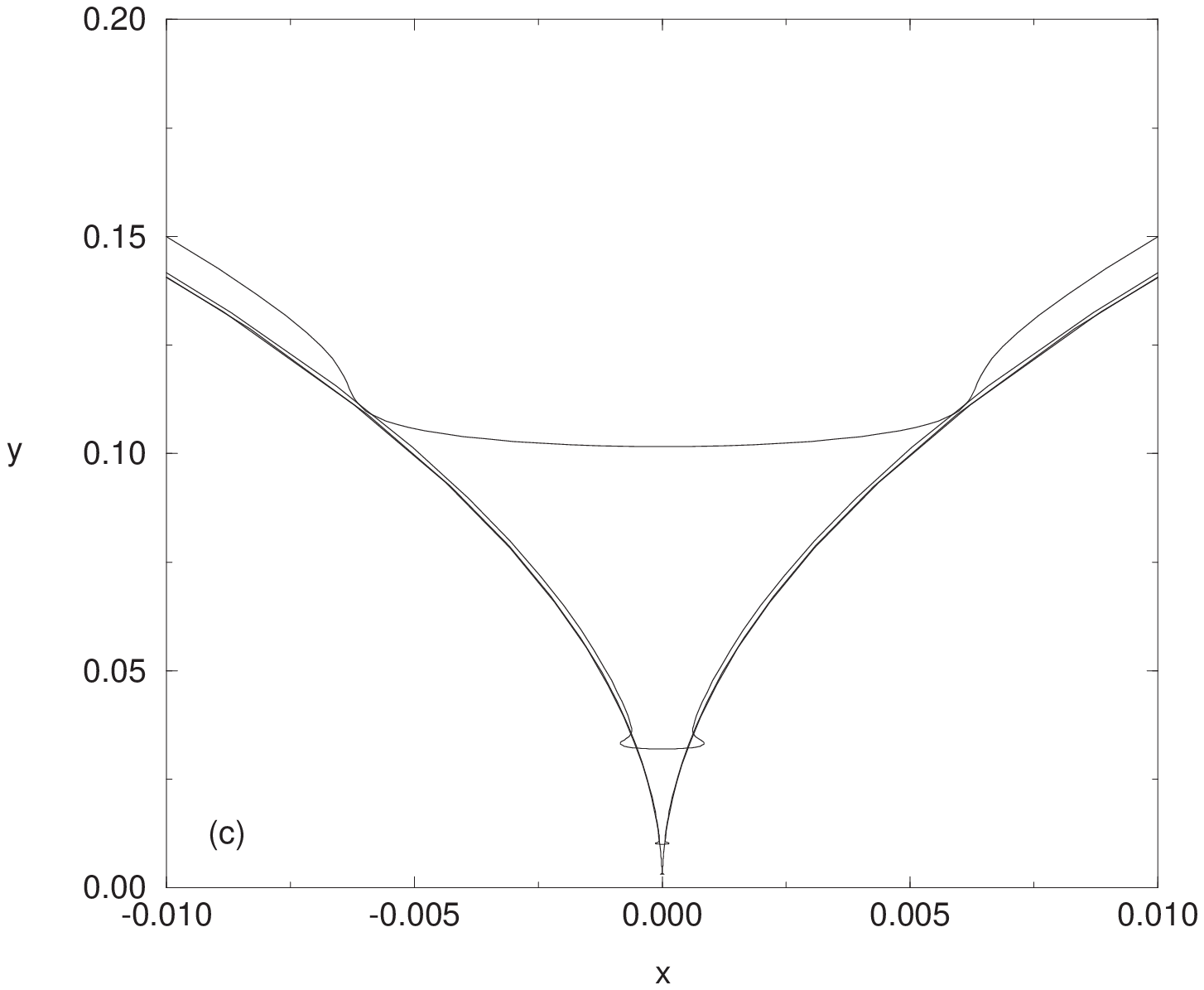}
  \end{center}
\end{figure}

\newpage
\begin{figure}
  {\Huge Figure 3:}
  \begin{center}
    \leavevmode
    \epsfsize=0.9 \textwidth
    \epsffile{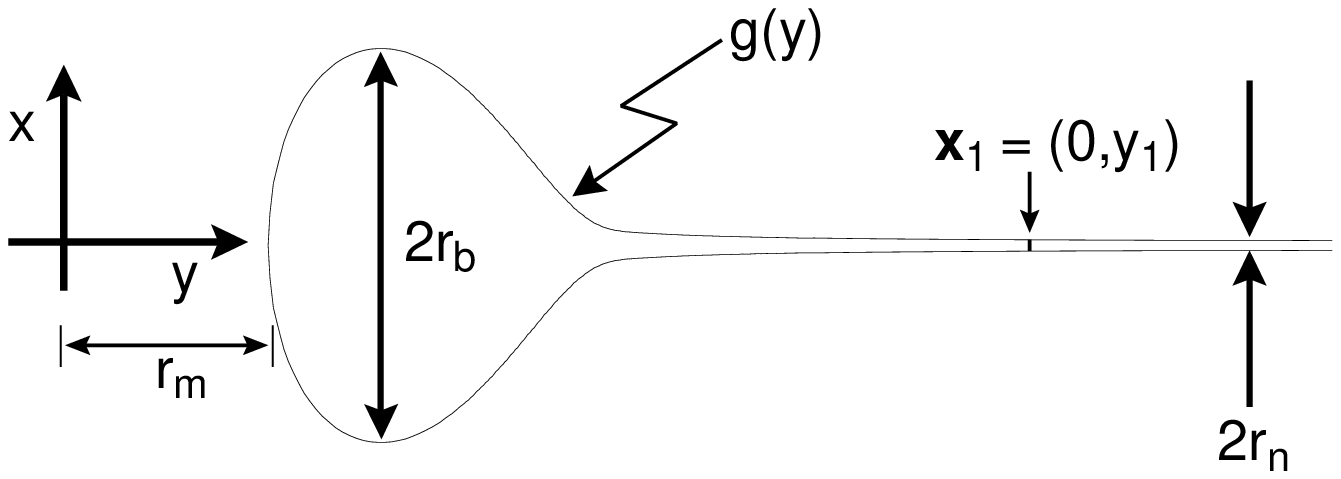}
  \end{center}
\end{figure}

\newpage
\begin{figure}
  {\Huge Figure 4:}
  \begin{center}
    \leavevmode
    \epsfsize=0.9 \textwidth
    \epsffile{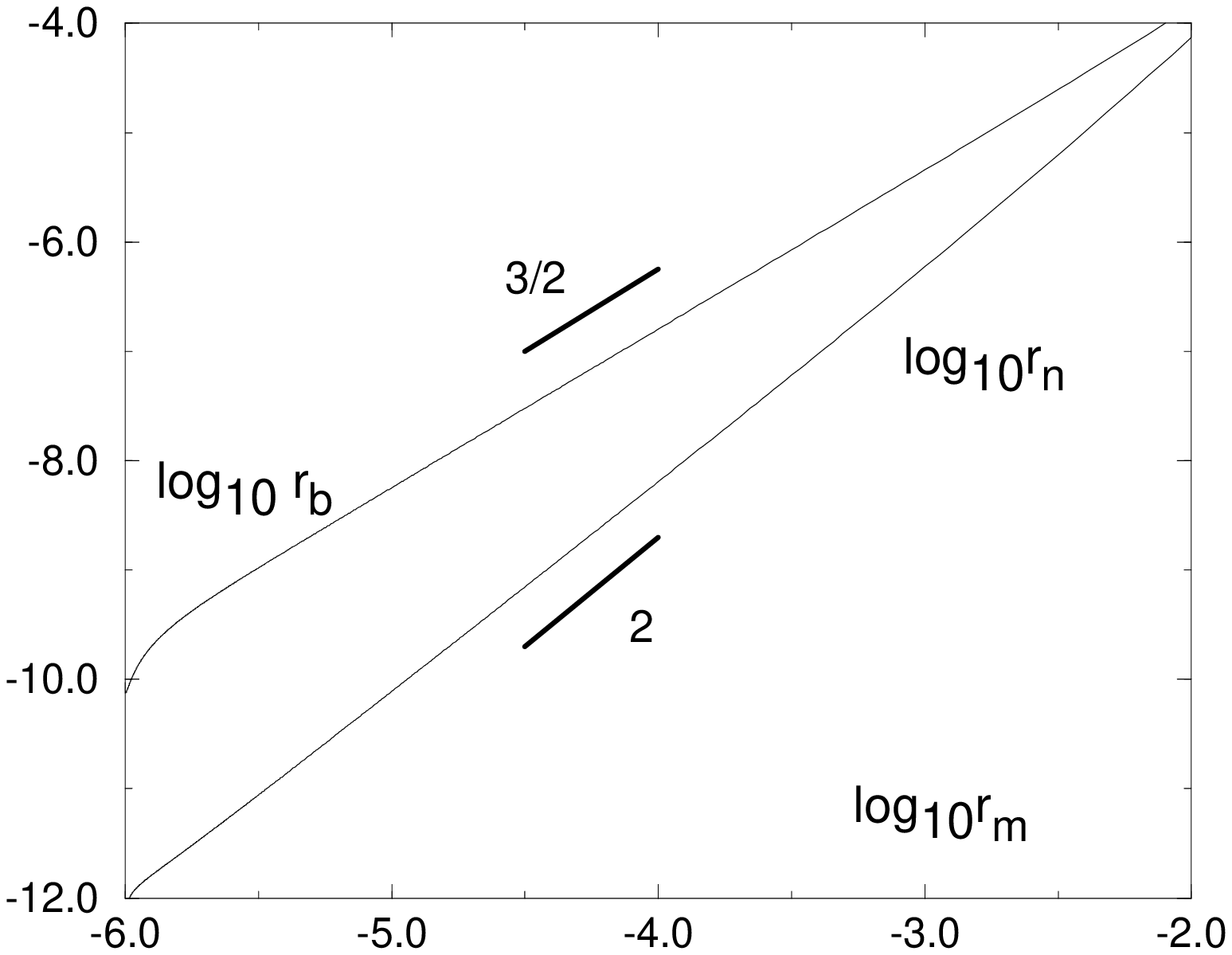}
  \end{center}
\end{figure}

\newpage
\begin{figure}
  {\Huge Figure 5:}
  \begin{center}
    \leavevmode
    \epsfsize=0.9 \textwidth
    \epsffile{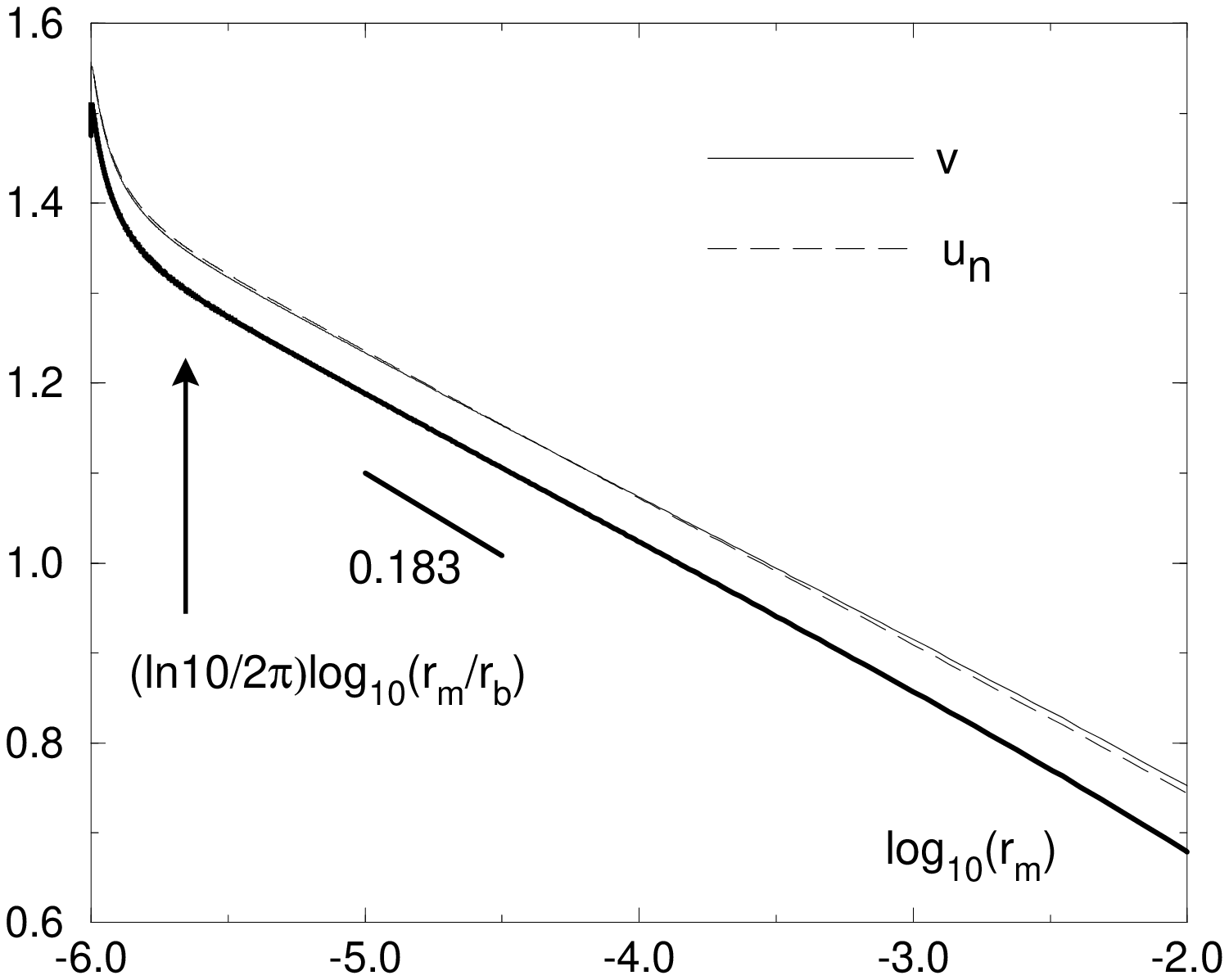}
  \end{center}
\end{figure}

\newpage
\begin{figure}
  {\Huge Figure 6:}
  \begin{center}
    \leavevmode
    \epsfsize=0.9 \textwidth
    \epsffile{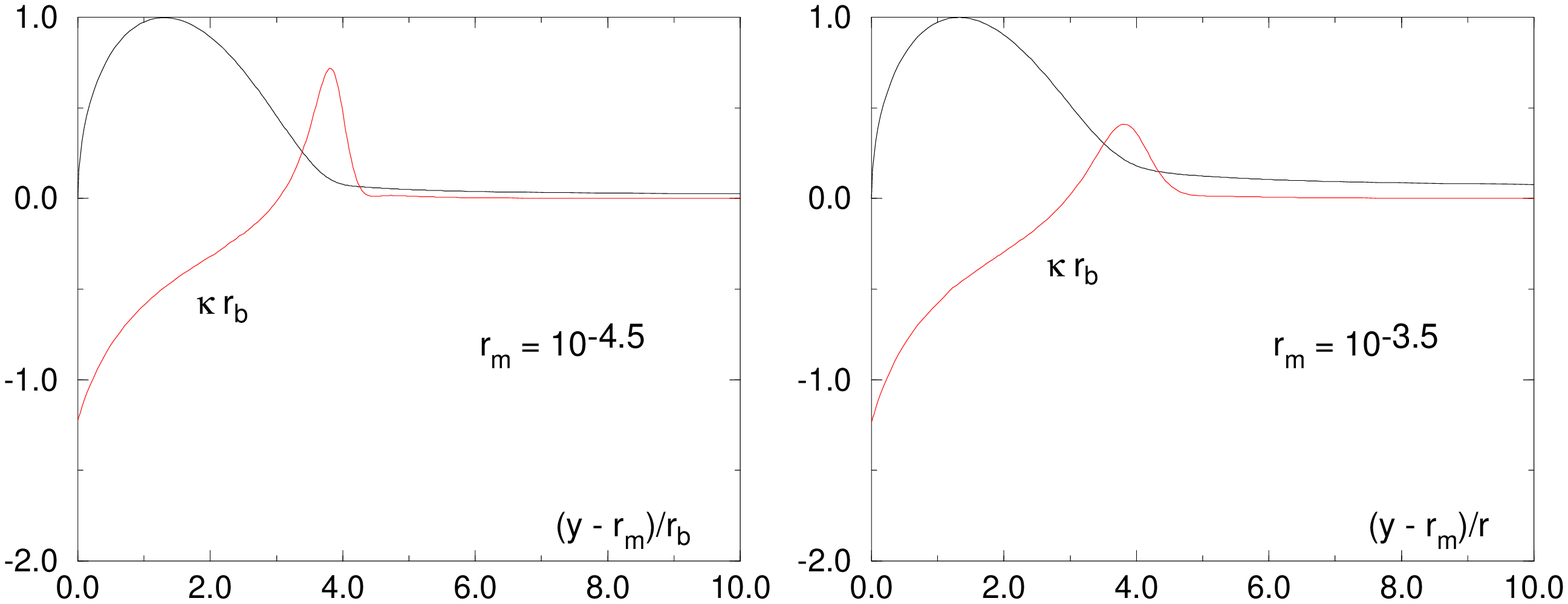}
  \end{center}
\end{figure}

\newpage
\begin{figure}
  {\Huge Figure 7:}
  \begin{center}
    \leavevmode
    \epsfsize=0.9 \textwidth
    \epsffile{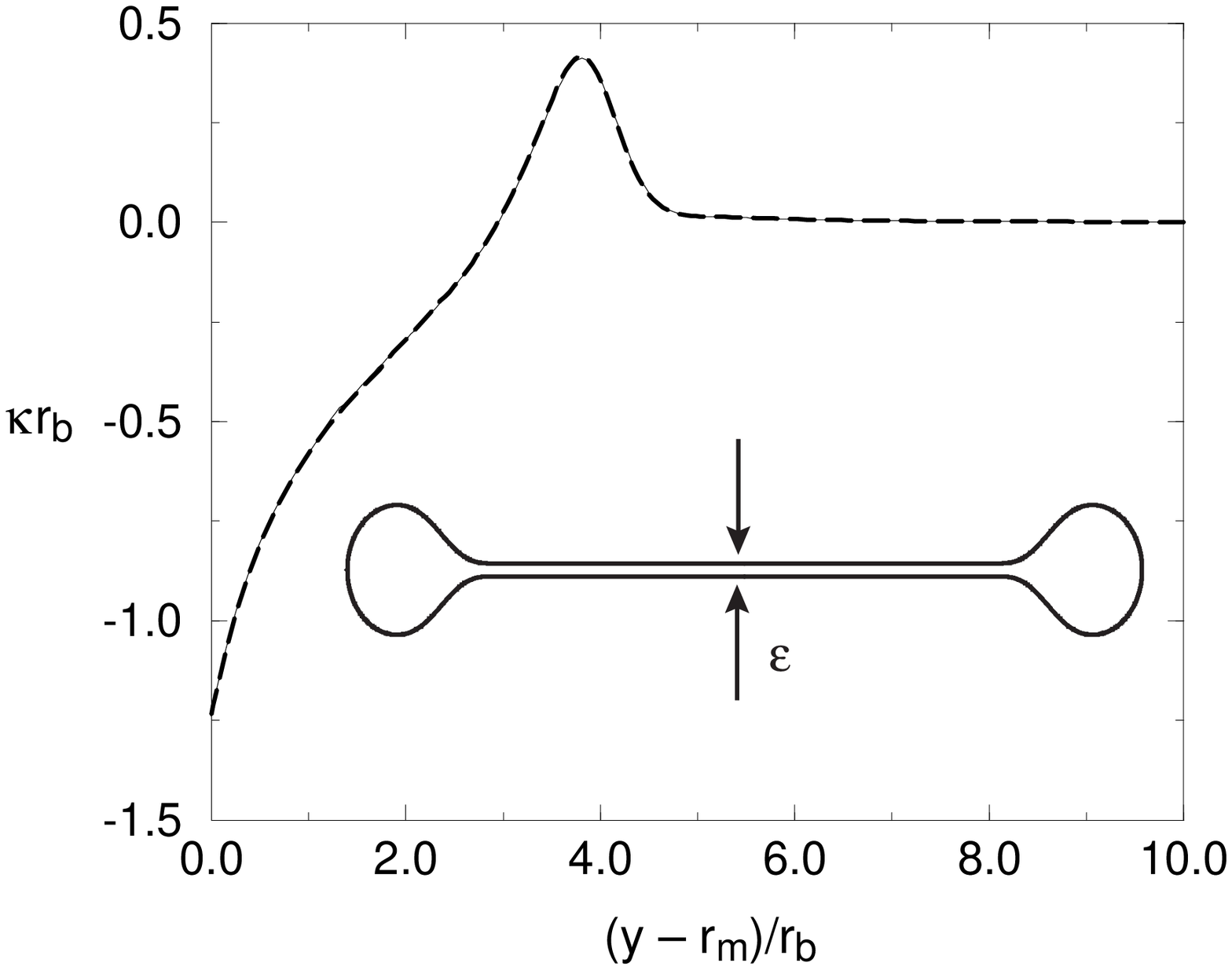}
  \end{center}
\end{figure}

\newpage
\begin{figure}
  {\Huge Figure 8a:}
  \begin{center}
    \leavevmode
    \epsfsize=0.95 \textwidth
    \epsffile{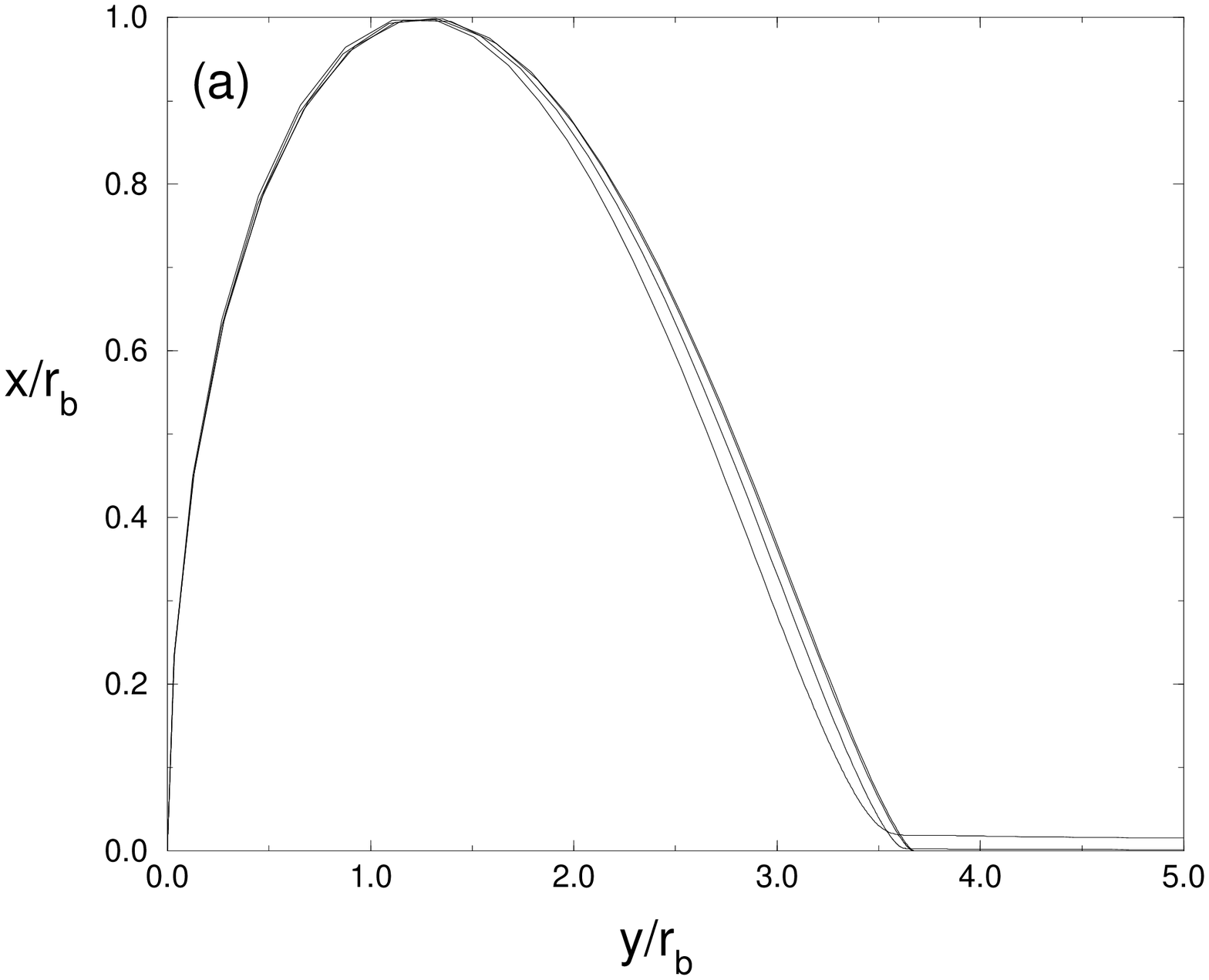}
  \end{center}
\end{figure}
\newpage
\begin{figure}
  {\Huge Figure 8b:}
  \begin{center}
    \leavevmode
    \epsfsize=0.95 \textwidth
    \epsffile{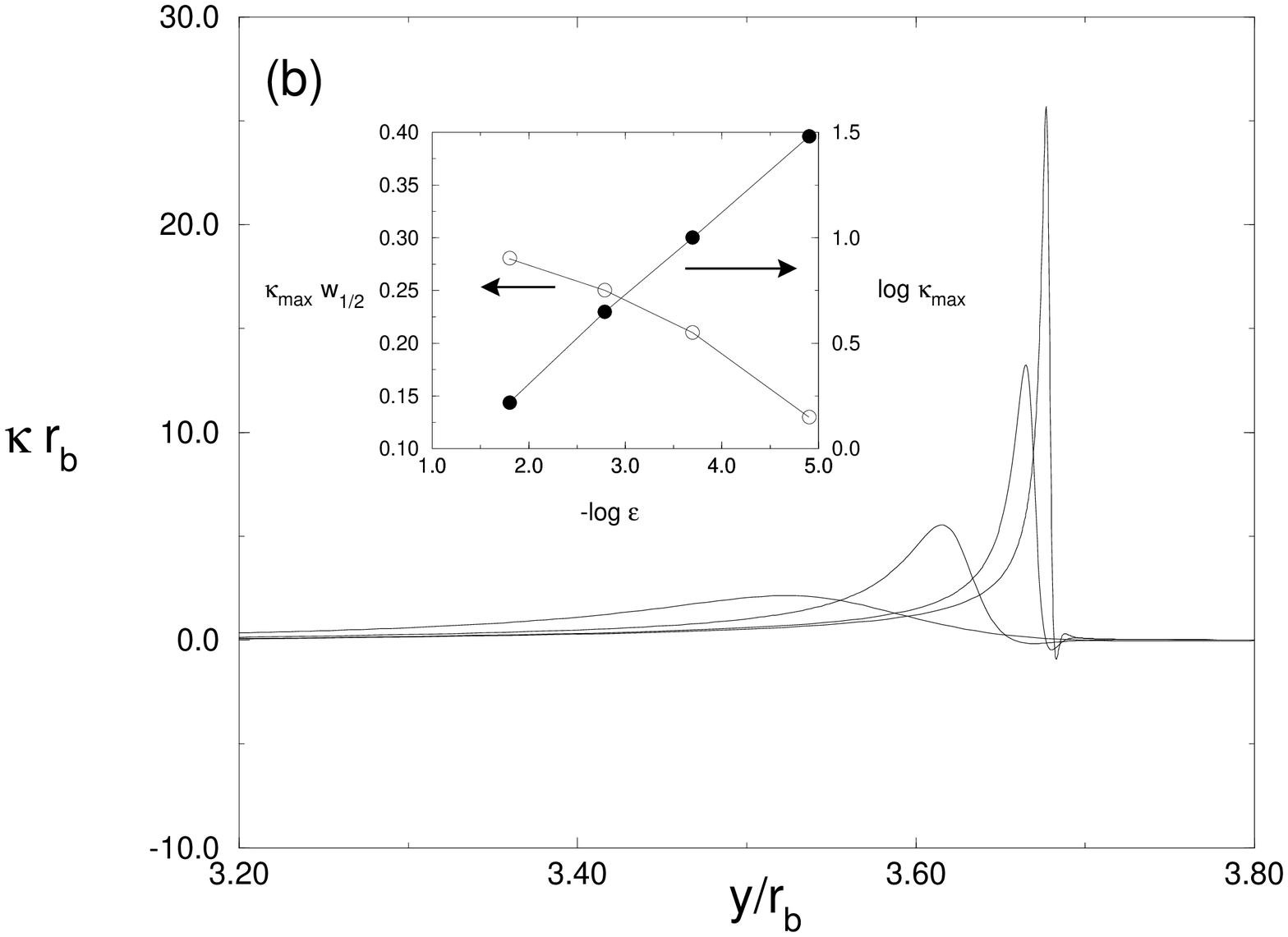}
  \end{center}
\end{figure}

\newpage
\begin{figure}
  {\Huge Figure 9:}
  \begin{center}
    \leavevmode
    \epsfsize=0.9 \textwidth
    \epsffile{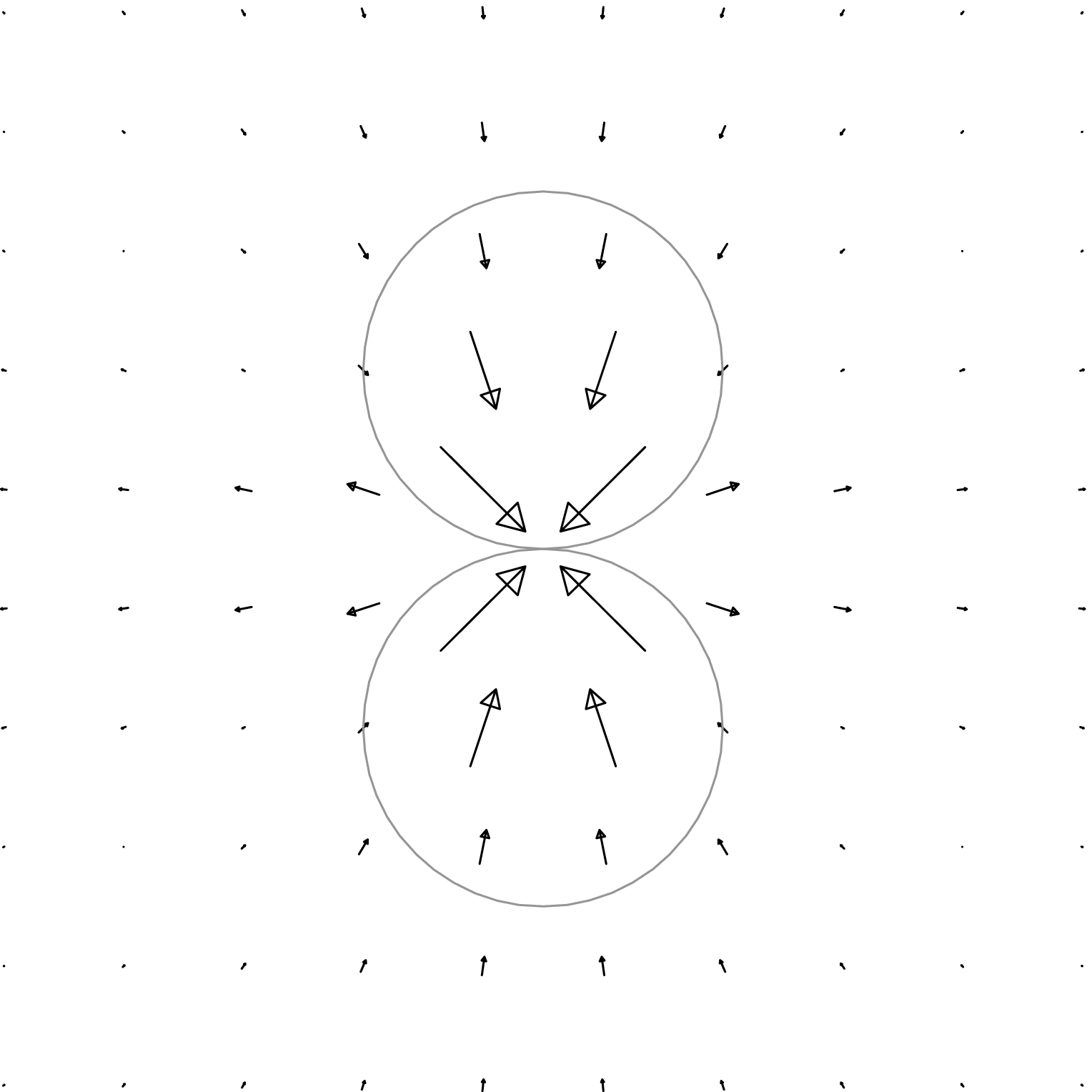}
  \end{center}
\end{figure}

\end{document}